\documentclass[showpacs,preprintnumbers,amsmath,amssymb]{revtex4}

\usepackage{epsfig}
\usepackage{axodraw}
\usepackage{graphicx}
\usepackage{dcolumn}% Align table columns on decimal point
\usepackage{bm}% bold math

\newcommand{\be}{\begin{equation}}
\newcommand{\ee}{\end{equation}}
\newcommand{\ba}{\begin{eqnarray}}
\newcommand{\ea}{\end{eqnarray}}
\newcommand{\jj}{J/\psi}
\begin{document}
\voffset2truecm
\preprint{BA-TH/2005-505, CERN-PH-TH/2005-009, FNT/T-2005/04}

\title{$J/\psi$ strong couplings to the vector mesons}
\author{V.~Laporta and A.D. Polosa}
\email{vincenzo.laporta@ba.infn.it; antonio.polosa@cern.ch}
\affiliation{
Dipartimento Interateneo di Fisica,
Universit\`a di Bari and INFN Bari, via Amendola 173, I-70126 Bari, Italy}
\author{F.~Piccinini}
\email{fulvio.piccinini@pv.infn.it}
\affiliation{Istituto Nazionale di Fisica Nucleare,  Sezione di Pavia and \\
Dipartimento di Fisica Nucleare e Teorica, via A. Bassi 6, I-27100, Pavia,
Italy}
\author{V.~Riquer}
\email{veronica.riquer@cern.ch}
\affiliation{CERN, Department of Physics, Theory Division, Geneva, Switzerland}
\date{\today}% It is always \today, today,

\begin{abstract}
We present a study of the cross sections $\jj\ X \to D^{(*)}\
\bar{D}^{(*)}$ ($X = \rho$, $\Phi$)  based on the calculation of the
effective tri- and four-linear couplings
 $\jj (X) D^{(*)} \bar{D}^{(*)}$
within a constituent quark model. In particular, the details of the
calculation of the four-linear couplings $\jj X D^{(*)}
\bar{D}^{(*)}$ are given. The results obtained have been used in a
recent analysis of $\jj$ absorption by the hot hadron gas formed in
peripheral heavy-ion collisions at SPS energies. 

\end{abstract}
\maketitle

\section{Introduction}

The problem of computing $\jj$ strong couplings to $\pi$, $\rho$ and
other pseudo-scalar and vector particles has its own interest
because it opens the way to the calculation of cross sections of the
kind (see Fig.~\ref{fig.absorption}):
\begin{equation}\label{eq.cross_section}
\sigma\left(\jj\ \{\pi,\rho,\ldots\}\ \longrightarrow\ D^{(*)}\ \bar
D^{(*)}\right).
\end{equation}
Such cross sections are the basic ingredients to estimate
the hadronic absorption background of $\jj$ in heavy-ion
collisions,  as it is thoroughly
discussed in~\cite{ioni,ioni2}.
The description of processes like those in Eq.~(\ref{eq.cross_section})
is a hard task because
no experimental test can be performed
and moreover they are not amenable to
first principles calculations, so that one has to resort to build models
and make approximations to describe their dynamics.

The dissociation process of the $\jj$ by hadrons has been considered
in several approaches, but the predicted cross sections show very
different energy dependence and magnitude near threshold.
Anyway, using different approaches, one consistently finds non negligible
cross section values (at least comparable with the nuclear one ${\cal N}\jj$, ${\cal N}=$nucleon)
especially for the reactions with $\pi$'s and $\rho$'s, the most studied
cases; for a review see for instance Ref.~\cite{barnes}.
This is certainly a clear indication that the picture of $\jj$
absorption by nuclear matter, as an antagonist mechanism to the
plasma suppression, is incomplete as long as interactions with the
hadronic gas formed in nucleus-nucleus collisions are not considered.

The problem of calculating the $\jj$ dissociation by pseudo-scalar
and vector mesons has been addressed in Refs.~\cite{ioni,ioni2}
within the \emph{Constituent-Quark--Meson} model (CQM),
originally devised to
compute exclusive heavy-light meson decays and tested on a quite
large number of such processes~\cite{cqm}. The basic calculations
refer to $\pi$ and $\rho$ contributions. The couplings to other mesons
have been obtained under the hypothesis of
flavour/octet symmetry.

The typical effective Feynman diagrams contributing to the $\jj$
dissociation are depicted in Fig.~\ref{fig.absorption}. The
tri-linear couplings $\rho D^{(*)}\bar{D}^{(*)}$ have been
calculated in~\cite{rhoa1} and the $\jj D^{(*)}\bar{D}^{(*)}$
couplings have been recently discussed in~\cite{pioni} (we report
these results at the end of Sect.~II),  where also  four-linear
couplings involving pions have been derived. The aim of the present
note is to explain the method used and the results obtained in
evaluating the four-linear couplings of the kind $\jj \rho D^{(*)}
\bar{D}^{(*)}$ (see Fig.~1, third diagram), since these are not
calculated elsewhere within the CQM framework. The numerical values
of the $\jj \Phi D_s^{(*)} \bar{D}_s^{(*)}$ couplings are also
given. For completeness we report also the expressions for the
tri-linear couplings discussed in Refs.~\cite{cqm,casalbuoni}. In
the end, we present the cross section predictions, based on the
complete set of contributing diagrams, for the processes $\jj\ \rho
\to D^{(*)}\ \bar{D}^{(*)}$ and $\jj\ \Phi \to D_s^{(*)}\
\bar{D}_s^{(*)}$, together with an estimate of the associated
theoretical uncertainties.

\section{The model}

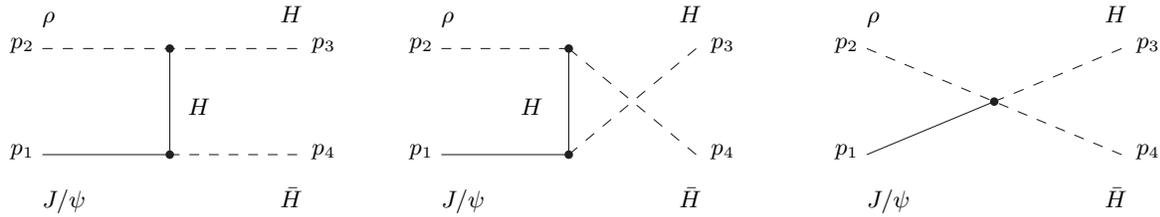
\begin{figure}
\begin{center}
\SetScale{0.8}
\noindent {\begin{picture}(300,50)(0,0)
\DashLine(0,50)(120,50){5} \Line(0,0)(60,0)
\DashLine(60,0)(120,0){5} \Vertex(60,50){2} \Vertex(60,0){2}
\Line(60,0)(60,50) \put(0,50){$\rho$} \put(0,-20){$J/\psi$}
\put(90,50){$H$} \put(90,-20){$\bar H$} \put(55,15){$H$}
\put(-12,0){$p_1$}
\put(-12,40){$p_2$}
\put(102,40){$p_3$}
\put(102,0){$p_4$}
\end{picture}
\hspace*{7em}
\begin{picture}(0,0)(220,0)
\DashLine(0,50)(60,50){5}
\DashLine(60,50)(120,0){5}
\Line(0,0)(60,0)
\DashLine(60,0)(120,50){5}
\Vertex(60,50){2}
\Vertex(60,0){2}
\Line(60,0)(60,50)
\put(0,50){$\rho$}
\put(0,-20){$J/\psi$}
\put(90,50){$H$}
\put(90,-20){$\bar H$}
\put(30,15){$H$}
\put(-12,0){$p_1$}
\put(-12,40){$p_2$}
\put(102,40){$p_3$}
\put(102,0){$p_4$}
\end{picture}
\hspace*{7em}
\begin{picture}(0,0)(130,0)
\DashLine(0,50)(120,0){5}
\Line(0,0)(60,25)
\DashLine(60,25)(120,50){5}
\Vertex(60,25){2}
\put(0,50){$\rho$}
\put(0,-20){$J/\psi$}
\put(90,50){$H$}
\put(90,-20){$\bar H$}
\put(-12,0){$p_1$}
\put(-12,40){$p_2$}
\put(102,40){$p_3$}
\put(102,0){$p_4$}
\end{picture}
}
\end{center}
\vskip 8pt \caption{Tree level effective Feynman diagrams for the
$\jj\ \rho\to H\bar H$ reaction, $H$ being $D^{(*)}$, with
$D^{(*)}=D$ or $D^*$.}\label{fig.absorption}
\end{figure}

\begin{figure}
\begin{center}
\includegraphics[scale=0.7]{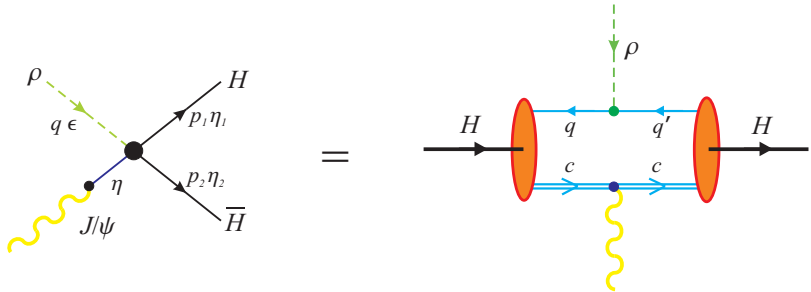}
\caption{Basic diagrammatic equation to compute the $g_4$ couplings.
The l.h.s. is the effective vertex $\jj\rho H\bar H$ at meson level
($\epsilon$ and $\eta$ are respectively the $\rho$ and $\jj$
polarizations); while the r.h.s. contains the 1-loop process to be
calculated in the CQM model.}\label{f:vertex}
\end{center}
\end{figure}

CQM is based on an effective Lagrangian which incorporates the
heavy quark spin-flavor symmetries and the chiral symmetry in the
light sector. In particular, it contains effective vertexes
between an heavy meson and its constituent quarks (see the
vertexes in the r.h.s. of Fig.~\ref{f:vertex}) whose emergence has
been shown to occur when applying bosonization techniques to
Nambu--Jona-Lasinio (NJL) interaction terms of heavy and light
quark fields~\cite{ebert}. On this basis we believe that CQM can
be considered as a quite reasonable approach to the computation of $\jj$
strong couplings to be compared to the various methods available in
the literature, often based on $SU(4)$ symmetry~\cite{altri}.

 In Fig.~\ref{f:vertex} we show
the typical diagrammatic equation to be solved in order to obtain
$g_4$($g_3$), four(tri)-linear couplings, in the various
cases at hand: on the l.h.s. it is represented the effective
four-linear coupling to be used in the cross section calculation
(to obtain one of the relevant  tri-linear couplings we could discard either the
$\jj$ or the $\rho$); the effective
interaction at the meson level (l.h.s.) is modeled as an
interaction at the quark-meson level (r.h.s. of Fig.~\ref{f:vertex}).

The $\jj$ is introduced using a Vector Meson Dominance~(VMD)
Ansatz: in the effective loop on the r.h.s. of Fig.~\ref{f:vertex}
we have a vector current insertion on the heavy quark line~$c$
while on the l.h.s. the $\jj$ is assumed to dominate the tower
of $1^-$, $c\bar c$ states mixing with the vector current (for
more details see~\cite{pioni}). Similarly, vector particles
coupled to the light quark component of the heavy mesons
$\rho,\omega$, when $q=(u,d)$, or $K^*,\Phi$, when one or both
light quarks involved are $q=s$, are also taken into account using
VMD arguments.

The pion and other pseudo-scalar fields have a
derivative coupling to the light quarks of the Georgi-Manohar
kind~\cite{manogeo}.

In this paper we will mainly focus on the reaction:
\begin{equation}\label{eq.reaction}
\jj\ \rho\ \longrightarrow\ D^{(*)}\ \bar{D}^{(*)}
\end{equation}
and in particular on the four-linear coupling $\jj\rho H\bar H$
(third graph) in  Fig.~\ref{fig.absorption}.

In CQM, as in \emph{Heavy Quark Effective Theory}
(HQET)~\cite{hqet}, the heavy super-field $H$ describes the
charmed states $D$ and $D^*$, respectively associated to the
annihilation operators $P_5,\ P^\mu$. $H$ is written in the
following way:
\begin{equation}\label{eq.field H}
H(v)=\frac{1+v\!\!\!/}{2}(P\!\!\!\!/-P_5\gamma_5),
\end{equation}
where $v$ is the four-velocity of the heavy meson; the limit of very
large heavy quark mass is understood. The heavy quark propagator is:
\begin{equation}
\frac{1+v\!\!\!/}{2}\frac{i}{v\cdot k},
\end{equation}
where $k$ is the residual momentum defined by the equation
$p^\mu_Q=m_Q v^\mu +k^\mu$ and related to the interaction of the
light degrees of freedom with the heavy quark ($k\sim {\cal
O}(\Lambda_{\rm QCD})$).

The $\rho$ is described by the interpolating field
$\rho^\mu$~\cite{casalbuoni} and its kinetic term in the effective
Lagrangian is built out by the tensor field strength:
\begin{equation}\label{eq.F_rho}
{\cal
F}_{\mu\nu}(\rho)=\partial_\mu\rho_\nu-\partial_\nu\rho_\mu+[\rho_\mu,
\rho_\nu].
\end{equation}
In the approach followed by~\cite{casalbuoni}, $\rho^\mu$ is
defined by $\rho^\mu=im_\rho/f_\rho \hat{\rho}^\mu$ where
$\hat{\rho}$ is a $3\times 3$ hermitian traceless matrix analogous
to the $3\times 3$ $\pi$ matrix of the pseudo-scalar octet.

In the following we will use the Feynman rules defined in~\cite{cqm}. The
interaction terms relevant to this calculation are:
\begin{equation}
-\bar q~\bar H~Q_v+\textrm{h.c.},
\end{equation}
which describes the vertex light quark ($q$), heavy quark
($Q_v$), heavy meson ($H$), and
\begin{equation}
\bar
q\left(\frac{m_\rho^2}{f_\rho}\,\epsilon^\mu\gamma_\mu\right)q,
\end{equation}
describing the vertex light quark, light quark, $\rho$. Here the
decay constant $f_\rho$ is defined by
\begin{equation}
\langle 0 |V_\mu|\rho(q,\epsilon)\rangle=if_\rho\epsilon_\mu,
\end{equation}
where $f_\rho=0.152$ GeV$^2$. We use a similar definition for the decay
constant of $\jj$ but we factor out the mass $m_J$ in the latter case,
thus obtaining $f_J=0.405$ GeV.

Once established the form of the effective vertexes occurring in
the loop diagram in the r.h.s. of Fig.~\ref{f:vertex}, one has
just to compute it using some regularization; we will adopt the Schwinger
proper time.

CQM does not include any confining potential and an
infrared cutoff $\mu$ is needed to prevent low
integration momenta to access the energy region where
confinement should be at work.
The kinematic condition for the free dissociation of $H$ in $m_Q$ and $m_q$ is
given by:
\begin{equation}
m_H>m_Q+m_q
\end{equation}
with $p_H=m_H\,v\approx m_Q\,v+k$. It follows that
\begin{equation}
k\cdot v>m_q.
\end{equation}
In the hadron rest frame we have $k^0>m_q$ and we can therefore require:
\begin{equation}
\mu\approx m_q.
\end{equation}

The value of the constituent light quark mass in the model at hand is
given by a gap equation~\cite{cqm}:
\begin{equation}
m_q-m_0-8\,G\,I_1(m_q^2)=0,
\end{equation}
where $G=5.25$ GeV$^{-2}$, $m_0$ is the current mass and the $I_1$
integral is defined in the Appendix. As a consistency check, putting a
zero current mass for the $u,d$ species we get a constituent
mass of 300 MeV, while for
a strange current mass of $m_0=131$~MeV we obtain a
strange constituent mass of 500~MeV using $\mu=300,500$~MeV
respectively in the calculation of $I_1$.

The residual momentum has an upper limit
given by the chiral symmetry braking scale $\Lambda_\chi\simeq 4\pi
f_\pi$ which we adopt as a UV cutoff~\cite{cqm}.

Then the momenta running in the loop are limited by two cutoff's:
$\mu$ and $\Lambda$. These two cutoff's are implemented by the
Schwinger regularization on the light propagator as
follows:
\begin{equation}
\int d^4l\frac{1}{(l^2-m_q^2)} \longrightarrow\int
d^4l\int_{1/{\Lambda^2}}^{1/{\mu^2}}ds\
e^{-s\left(l^2+m_q^2\right)}.
\end{equation}

The diagrammatic equation in Fig.~\ref{f:vertex} states
that the effective vertex $\jj\rho H\bar H$ is given by:
\begin{eqnarray}\label{eq.VMD}
&&(-1)\sqrt{Z_H m_H Z_{H^\prime} m_{H^\prime}}\times\nonumber
\\
&&\times\,
N_c\int\frac{d^4l}{(2\pi)^4}\textrm{Tr}\left[\left(-i\bar
H^\prime(v')\right)\frac{i}{v'\cdot l+\Delta}\
\frac{m^2_J}{f_Jm_J}\eta\!\!\!/\ \frac{i}{v\cdot l+\Delta}\
\left(-iH(v)\right)\ \frac{i}{l\!\!/ -m}\
i\frac{m_\rho^2}{f_\rho}\epsilon\!\!/\ \frac{i}{l\!\!/+q\!\!\!/
-m}\right].~~~~~~
\end{eqnarray}
$H$ and $\bar H^\prime$ represent the heavy-light external meson
fields labeled by their four-velocities $v,v^\prime$ while
the
$\sqrt{Z_H m_H Z_{H^\prime} m_{H^\prime}}$ coupling factor
of heavy mesons to quarks is calculated in~\cite{cqm}.
The parameter $\Delta$ appearing in the
heavy propagator is defined by:
\begin{equation}
\Delta=M_H-m_Q,
\end{equation}
i.e., the mass of the heavy-light meson minus the mass of the heavy
quark contained in it. $\Delta$ is the main free parameter of
the model. It varies in the range $\Delta=0.3,0.4,0.5$~GeV for
$u,d$ light quarks and $0.5,0.6,0.7$~GeV for strange
quarks~\cite{eff0}. Varying $\Delta$ gives an handle to estimate the
theoretical error. $m$ is the constituent mass of light quarks as
defined above.

\section{The calculation}

The $\rho$ is coupled to the light quarks by VMD,
$\epsilon$ being its polarization and $q$ its 4-momentum. The
$\jj$, having polarization $\eta$, is also coupled via VMD, but to
the heavy quarks ($\eta$ appears in the trace between the two
heavy quark propagators, while $\epsilon$ appears between the two
light quark propagators). In front of this expression we have the
fermion loop factor.

The trace computation in~(\ref{eq.VMD}) will introduce a number of
scalar combinations of the momenta and polarizations of the
external particles that we will list in the Appendix.
Each of these combinations
will be weighted by a scalar integral which amounts to a numerical
factor: what we call the coupling. Actually, as we will see, such
scalar integrals depend on the energy of the $\rho$. In general
the expressions obtained for the four-linear couplings appear to
be quite complicated functions of $E_\rho$ if compared, e.g., to those
obtained when studying only $\jj$ interactions with
pions~\cite{pioni}. It is therefore difficult to write
down general polar expressions for the $E_\rho$ behaviour.
On the other hand we have in mind to use these
results to compute cross sections $\sigma_{\jj\rho}$ and thermal
averages $\langle \rho\cdot\sigma\rangle_T$ in a hadron gas at a
temperature $T\approx 170$~MeV where the Boltzmann factor is
presumably more effective than any polar form factor in damping
the high energy tails.

Using the Feynman trick the fermion loop of the above equation
(\ref{eq.VMD}) becomes:
\begin{equation}\label{eq.tr}
\frac{m^2_J}{f_Jm_J}\,\frac{m_\rho^2}{f_\rho}\,\sqrt{Z_Hm_HZ_{H'}m_{H'}}
\int_0^1dx\frac{\partial}{\partial{m^2(x)}}iN_c\int\frac{d^4l}{(2\pi)^4}
\frac{\textrm{Tr}\left[\bar
H'~\eta\!\!\!/~H~(l\!\!/-q\!\!\!/x+m)~\epsilon\!\!/~(l\!\!/-q\!\!\!/x+q\!\!\!/+m)\right]}
{(l^2-\tilde m^2)~(v\cdot l+\delta)~(v'\cdot l+\delta')} ,
\end{equation}
in which we have defined:
\begin{equation}\label{eq.m}
\tilde m^2=m^2+x\,m_\rho^2\,(x-1),
\end{equation}
and
\begin{eqnarray}
\delta&=&\Delta-x\,q\cdot v=\Delta-x\,E_\rho\label{eq.delta}
\\
\delta'&=&\Delta-x\,q\cdot v'=\Delta-x\,\omega\,
E_\rho,\label{eq.deltapr}
\end{eqnarray}
where $E_\rho$ is the energy of the incident $\rho$ and
$\omega=v\cdot v'$ ($v^\prime=\omega v$).
The cross section computation is performed in
the frame where $\jj$ is at rest. $\omega$, in this frame, is
related to the meson masses by:
\begin{equation}\label{eq.omega}
\omega=\frac{m_{\jj}^2+m_\rho^2-m_H^2-m_{H^\prime}^2+2E_\rho
m_{\jj}}{2m_Hm_{H^\prime}}.
\end{equation}

By kinematic considerations the energy threshold of the reactions
(\ref{eq.reaction}) for $D\bar D$ and $D^*\bar D$ channels is
$E_\rho\simeq 0.77$ GeV whereas for $D^*\bar D^*$ channel is
$E_\rho\simeq 0.96$ GeV, with $\omega\approx 1$. We
consider $\rho$ particles with energies in the range between
$0.77$ and $1$~GeV where the two final state mesons are almost at rest.

All the couplings that we can extract by direct computation can be
written in terms of 7 basic expressions which we call:
$L_5,A,B,C,D,E,F$. The latter are linear combinations of the
$I_i,L_i$ integrals listed in the Appendix and are defined by:
\begin{equation}
\frac{\partial}{\partial{\tilde{m}^2}}iN_c\int\frac{d^4l}{(2\pi)^4}
\frac{1}{(l^2-\tilde m^2)~(v\cdot l+\delta)~(v'\cdot
l+\delta')}=L_5
\end{equation}

\begin{equation}
\frac{\partial}{\partial{\tilde{m}^2}}iN_c\int\frac{d^4l}{(2\pi)^4}
\frac{l_\mu}{(l^2-\tilde m^2)~(v\cdot l+\delta)~(v'\cdot
l+\delta')}=A\,v_\mu+B\,v_\mu'
\end{equation}

\begin{equation}
\frac{\partial}{\partial{\tilde{m}^2}}iN_c\int\frac{d^4l}{(2\pi)^4}
\frac{l_\mu l_\nu}{(l^2-\tilde m^2)~(v\cdot l+\delta)~(v'\cdot
l+\delta')}=C\,g_{\mu\nu}+D\,v_\mu v_\nu+E\,v'_\mu
v'_\nu+F\,(v_\mu v'_\nu+v'_\mu v_\nu).
\end{equation}
The final
expression of the loop integral can the be reduced to a sum of terms of
the general form:
\begin{equation}
\sum_SS(H,H')\ C\int_0^1dx~g_4^{(S)}(x,E_\rho)
\end{equation}
where $S(H,H')$ represent the scalar combinations of
momenta and polarizations of $H$ and $H'$ occurring in the calculation;
$g_4^{(S)}$, are the
corresponding couplings. Here $C$ is given by:
\begin{equation}
C=\frac{m^2_J}{f_Jm_J}\,\frac{m_\rho^2}{f_\rho}\,\sqrt{Z_Hm_HZ_{H'}m_{H'}},
\end{equation}
with $f_J=0.405$ GeV. Our central values for the couplings are obtained
for $\Delta=0.4$~GeV (and
$Z_H=2.36$ GeV$^{-1}$~\cite{cqm}), while for $m_H$ we use the
experimental value for the $D^{(*)}$ mass.

In Table \ref{tab.formfactorsJrhoHH} we list the explicit
expressions of the couplings. We call them $g_4=\{g,h,f\}$,
respectively related to the four linear couplings $\jj\rho D\bar
D$, $\jj\rho D^*\bar D $, $\jj\rho D^*\bar D^*$. All these
expressions have to multiplied by $C$ and integrated over $x$; the
numerical values in Table I are then:
\begin{equation}
C\int_0^1dx~g_4(x,E_\rho).
\end{equation}
These are typically complicated functions of $E_\rho$ and it is
not as easy as in~\cite{pioni} to determine an explicit polar form
factor dependency common to all of them. On the other hand we have in
mind to adopt these couplings to compute cross sections of the
kind $\sigma_{\jj\rho\,\to\,{\rm open~charm}}$ and use this
information to compute thermal averages in the form:
\begin{equation}
\left\langle \rho\cdot\sigma_{\jj\rho\,\to\,{\rm
open~charm}}\right\rangle_T = \frac{N}{2\pi^2}\int_{E_0}^{\infty}
dE\ \frac{p\,E\,\sigma(E)}{e^{E/kT}-1}, \label{eq.thave}
\end{equation}
where $E_0$ is the energy threshold needed to open the reaction
channel and the Bose distribution is used to describe an ideal gas
of mesons. $\rho$ in the l.h.s of Eq.~(\ref{eq.thave}) is the number
density of particles in the gas. The Boltzmann factor $\exp\,(-E/T)$
will be at work as an exponential form factor cutting high energy
tails faster than any polar one. We could therefore avoid any
arbitrary Ansatz on form factors at the interaction vertexes. We
limit ourself to study the dependency of our couplings on $E_\rho$
in the range of energy where we reasonably think to have $\rho$
mesons in the hadron gas excited by a peripheral heavy-ion
collision. Estimating the $\jj$ absorption background to the
suppression signal in heavy-ion collisions amounts to compute the
attenuation lengths (inverse of the thermal averages in
Eq.~(\ref{eq.thave})) in a hot gas, $T\approx 170$~MeV, populated by
$\pi,K,\eta,\rho,\omega,...$ mesons. We could therefore expect to
have $E_\rho\approx 770\div 1000$~MeV.

\begin{table}[h!]
\begin{displaymath}
\begin{array}{ccrclrclc}
\hline\hline
~~~~~~~~~~&\phantom{\Big(}\jj XD\bar D&\multicolumn{3}{c}{X=\rho}&\multicolumn{3}{c}{X=\Phi}&~~~~~~~~~~~~~~~~\\
\hline &&&&&&&&
\\
g_1&\frac{2}{m_D^3}Ax&4&\pm&2&1.5&\pm&0.5&\textrm{GeV}^{-4}\\
g_2&\frac{2}{m_D^3}B(x-1)&-2.3&\pm&1.0&-1.1&\pm&0.2&\textrm{GeV}^{-4}\\
g_3&\frac{1}{m_D}\left(A+B+2xA(\omega-1)-mL_5\right)&27&\pm&4&13&\pm&1&\textrm{GeV}^{-2}\\
g_4&\frac{2}{m_D^2}(D+F-Am)&-9&\pm&3&-7&\pm&1&\textrm{GeV}^{-2}\\
g_5&\frac{1}{m_D^2}((m^2+m_\rho^2x(1-x))L_5-2Am-2C+D-E+2F(1-\omega))&-8&\pm&3&-7&\pm&1&\textrm{GeV}^{-2}\\
g_6&\frac{1}{m_D}(A-B+2B(x-\omega x+\omega)-mL_5)&25&\pm&4&12&\pm&1&\textrm{GeV}^{-2}\\
g_7&\frac{1}{m_D^2}((m^2+m_\rho^2x(1-x))L_5-2Bm-2C-D+E+2F(1-\omega))&-6&\pm&2&-5&\pm&1&\textrm{GeV}^{-2}\\
g_8&\frac{2}{m_D^2}(E+F-Bm)&-7&\pm&2&-5&\pm&1&\textrm{GeV}^{-2}\\
g_9&((m^2+m_\rho^2x(1-x))L_5-2C-D-E-2F\omega)(1-\omega)&-0.4&\pm&0.4&-0.4&\pm&0.2&\\
&&&&&&&&
\\
\hline \hline
&\phantom{\Big(}\jj XD^*\bar D&\multicolumn{3}{c}{X=\rho}&\multicolumn{3}{c}{X=\Phi}&~~~~~~~~~~~~~~~~\\
\hline &&&&&&&&
\\
h_1&(mL_5+(A-B)x)(\omega-1)&1&\pm&2&0.1&\pm&0.6&\textrm{GeV}^{-1}\\
h_2&B(x-1)&-9&\pm&4&-5&\pm&1&\textrm{GeV}^{-1}\\
h_3&mL_5-Bx&-6&\pm&12&-6&\pm&3&\textrm{GeV}^{-1}\\
h_4&mL_5+A(x-1)-B&-35&\pm&15&-20&\pm&4&\textrm{GeV}^{-1}\\
h_5&A&35&\pm&11&15.7&\pm&3&\textrm{GeV}^{-1}\\
h_6&Ax&15&\pm&8&6&\pm&2&\textrm{GeV}^{-1}\\
h_7&B-mL_5&16&\pm&16&11&\pm&4&\textrm{GeV}^{-1}\\
h_8&(m^2+m_\rho^2x(1-x))L_5-2C-D-E-2F\omega&1.3&\pm&2&1.1&\pm&0.8&\\
h_9&D+F-mA&-19&\pm&7&-17&\pm&3&\\
h_{10}&E+F-mB&-15&\pm&6&-13&\pm&2&\\
&&&&&&&&
\\
\hline \hline &\phantom{\Big(}\jj XD^*\bar
D^*&\multicolumn{3}{c}{X=\rho}&\multicolumn{3}{c}{X=\Phi}&~~~~~~~~~~~~~~~~\\
\hline &&&&&&&&
\\
f_1&\frac{1}{m_{D^*}}(A+B-mL_5)&25.5&\pm&0.6&13.6&\pm&0.1&\textrm{GeV}^{-2}\\
f_2&\frac{1}{m_{D^*}}(B-mL_5+A(2\omega x-2x+1))&26&\pm&1&13.86&\pm&0.08&\textrm{GeV}^{-2}\\
f_3&\frac{2}{m^3_{D^*}}Ax&4&\pm&2&1.5&\pm&0.5&\textrm{GeV}^{-4}\\
f_4&\frac{1}{m_{D^*}}(A-mL_5+B(-2\omega x+2x+2\omega-1))&26.0&\pm&0.5&13.8&\pm&0.1&\textrm{GeV}^{-2}\\
f_5&\frac{2}{m^3_{D^*}}B(x-1)&-2.2&\pm&0.7&-1.2&\pm&0.1&\textrm{GeV}^{-4}\\
f_6&(-m^2L_5+2C+D+E+2F\omega+m_\rho^2(x^2-x)L_5)(1-\omega)&0.03&\pm&0.1&0.03&\pm&0.03&\\
f_7&\frac{1}{m^2_{D^*}}(-m^2L_5+2C+D+E+2F\omega+m_\rho^2(x^2-x)L_5)&-0.1&\pm&2&-0.2&\pm&0.2&\textrm{GeV}^{-2}\\
f_8&\frac{2}{m^2_{D^*}}(D+F-mA)&-8&\pm&2&-8.2&\pm&0.3&\textrm{GeV}^{-2}\\
f_9&\frac{1}{m^2_{D^*}}(-m^2L_5+2mA+2C-D+E+2F(\omega-1)+m_\rho^2(x^2-x)L_5)&8.3&\pm&2&8.0&\pm&0.1&\textrm{GeV}^{-2}\\
f_{10}&\frac{1}{m^2_{D^*}}(-m^2L_5+2mB+2C+D-E+2F(\omega-1)+m_\rho^2(x^2-x)L_5)&7.3&\pm&0.8&6.3&\pm&0.2&\textrm{GeV}^{-2}\\
f_{11}&\frac{2}{m^2_{D^*}}(E+F-mB)&-7&\pm&1&-6.5&\pm&0.5&\\
&&&&&&&&
\\
\hline \hline
\end{array}
\end{displaymath}
\caption{The couplings $g_4=\{\,g_i,h_i,f_i\,\}$ expressed as
linear combinations of the basic scalar integrals listed in the
Appendix. The numerical values are given by $C\int_0^1dx\,g_4$ :
the mean values are estimated by setting $\Delta=0.4$ GeV
($\Delta=0.6$ GeV) and varying the energy of $\rho$ ($\Phi$) in
the range $E_\rho=0.770-1$~GeV, ($E_\Phi=1.02-1.2$ GeV), while the
error is calculated by combining the excursion in the selected energy range
at $\Delta=0.4$ GeV with the uncertainty over $\Delta$  (in the range
$\Delta=0.5-0.7$~GeV). Some of the
couplings are unavoidably affected by large uncertainties. This
affects the determination of the attenuation lengths discussed
in the text to the extent
pointed out in~\cite{ioni2}.}\label{tab.formfactorsJrhoHH}
\end{table}

The loop that must be calculated in the case in which we substitute a $\Phi$
particle to the $\rho$ is the same as in
Fig.~\ref{f:vertex} but with $q,q'=s$ ($m_s=500$~MeV, $\mu\sim m_s$),
and with super-fields $H_s$ in place of $H$.
The reaction in this case is
\begin{equation}
\jj\ \Phi\ \longrightarrow\ D_s^{(*)}\ \bar{D}_s^{(*)}.
\end{equation}

The structure of the coupling of $\Phi$ to light quark current is
identical to the one of the $\rho$,
and all the above equations are valid also in this case
with the substitutions: $m_\rho\to m_\Phi$, $f_\rho\to f_\Phi$,
$H\to H_s$. Numerically we have used $f_\Phi=0.249$ GeV$^2$, while
for $m_{H_s}$ we have used the experimental value for $D_s^{(*)}$.
The numerical values are reported in Table \ref{tab.formfactorsJrhoHH}.
\\

We conclude this section by writing the tri-linear couplings
$\rho HH$; they are computable within the same framework by simply
not including the $\jj$ in the diagrammatic equation of
Fig.~\ref{f:vertex}.

The vertex $\rho D^{(*)}\bar D^{(*)}$ is described by two constants
$g_3=\beta,\lambda$ and the effective Lagrangian describing this
interaction can be written as~\cite{casalbuoni,cqm}:
\begin{equation}
{\cal L}_{HH\rho}=-i\beta\,\textrm{Tr}[\bar
HH]~v\cdot\rho+i\lambda\,\textrm{Tr}[\bar H\sigma_{\mu\nu}H]{\cal
F}^{\mu\nu},
\end{equation}
where the field $\rho^\mu$ and the tensor ${\cal F}^{\mu\nu}$ have
been defined above. The numerical values are:
\begin{eqnarray}
\beta&=&-0.98\\
\lambda&=&+0.42\ \textrm{GeV}^{-1};
\end{eqnarray}
analogously $\beta$ and $\lambda$ for the ${\cal L}_{HH\Phi}$ are
\begin{eqnarray}
\beta&=&-0.48\\
\lambda&=&+0.14\ \textrm{GeV}^{-1}.
\end{eqnarray}

As for the couplings $\jj D^{(*)}\bar D^{(*)}$, they have been
extensively discussed in~\cite{pioni}. Here we just report the main results.
Observe that
\begin{equation}
{\cal L}_{\jj HH}=ig_{\jj HH}\,\textrm{Tr}[\bar H\gamma_\mu
H]J^\mu,
\end{equation}
where $H$ can be any of the pairs $D\,D^*$ or $D_s\,D_s^*$
(neglecting $SU(3)$ breaking effects). As a consequence of the
spin symmetry of the HQET we find:
\begin{equation}
g_{\jj D^*D^*}=g_{\jj DD}\,,~~~~~~~~~~
g_{\jj DD^*}=\frac{g_{\jj DD}}{m_D}.
\end{equation}
The numerical values are given by:
\begin{eqnarray}
g_{\jj DD}&=&8.0\pm 0.5 \nonumber\\
g_{\jj DD^*}&=&4.05\pm 0.25{\ \rm GeV}^{-1}\nonumber\\
g_{\jj D^* D^*}&=&8.0\pm 0.5. \nonumber
\end{eqnarray}

\begin{figure}[h!]
\begin{center}
\includegraphics[scale=.8]{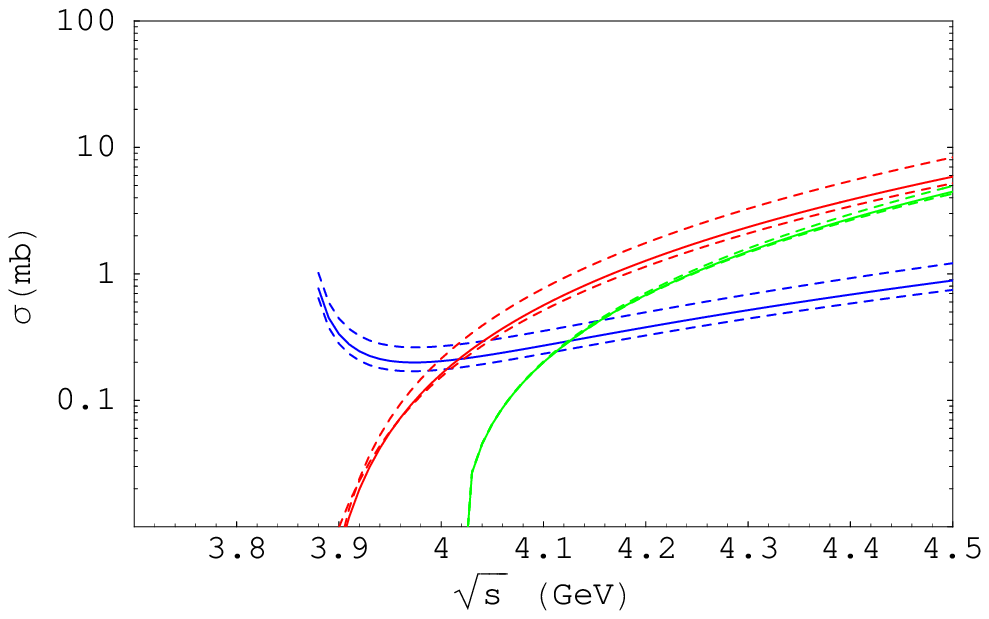}
\includegraphics[scale=.8]{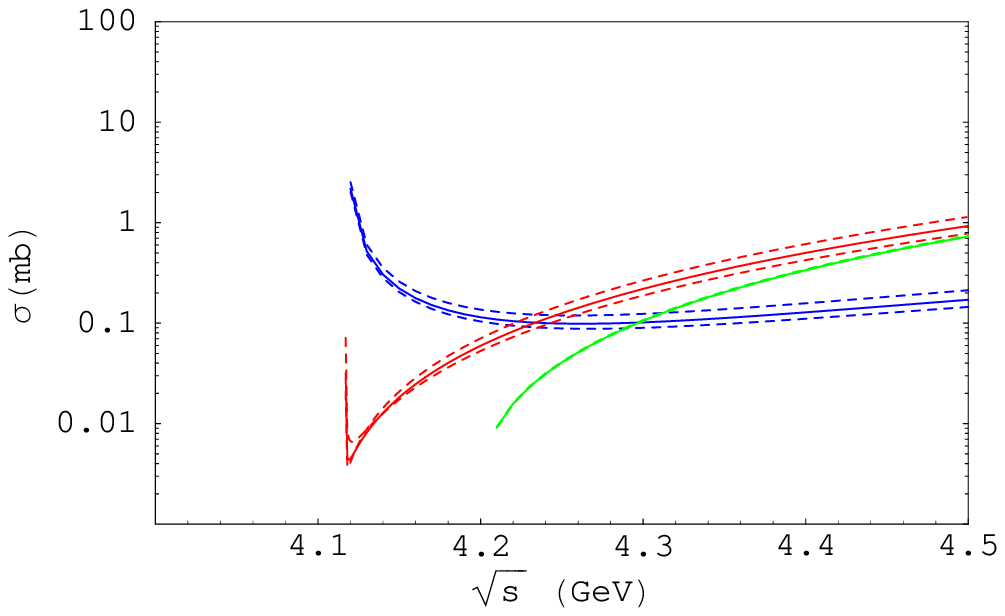}
\caption{The cross sections of the processes $\jj\
\rho\longrightarrow D^{(*)}\,\bar D^{(*)}$ and $\jj\
\Phi\longrightarrow D_s^{(*)}\,\bar D_s^{(*)}$ on the left and on
the right panel respectively. The dashed curves define the uncertainty
bands obtained by varying $\Delta$ and $E_\rho$ as discussed above.
Some of the reactions, the one initiated by $\rho$ giving $D\bar{D}$ in the final state and
those initiated by $\phi$ giving $D_{s}\bar{D}_{s}$ and $D^*_{s}\bar{D}_{s}$ 
(or  $D_{s}\bar{D}^*_{s}$, a sum of the two is taken) show the typical ``exothermic'' peak for zero 
$\rho (\phi)$ momentum.
The remaining reactions show the usual threshold behaviour (endothermic).
}\label{f:tre}
\end{center}
\end{figure}

In Fig.~\ref{f:tre} we report the cross section curves as functions of
the $\sqrt{s}$ of the process
for the three final states under consideration $(DD,DD^*,D^*D^*)$.
This calculation has been made by using the tri- and four-linear couplings
quoted above, assuming their validity in the energy range $\sqrt{s}\approx 3.8\div 4.5$,
and computing the tree level diagrams
for the process at hand (for a sketch of the diagrams
involved see~\cite{ioni}). The dashed curves define the uncertainties
bands obtained by varying $\Delta$ and $E_\rho$, as discussed in
Tab~\ref{tab.formfactorsJrhoHH}.

\section{Summary}
We have presented the calculation method of the effective couplings
$\jj (X) D^{(*)} \bar{D}^{(*)}$, with $X = \rho$, $\Phi$, within
the CQM model.
The resulting cross section predictions,
together with an estimate of the associated theoretical uncertainties, have
been presented as functions of $\sqrt{s}$,
showing values of the same order as the cross sections
for $\jj \pi \to D^{(*)} \bar{D}^{(*)}$.
This, given also the higher spin multiplicity of the $\rho$ meson
with respect to pions, demonstrates the importance of the $\rho$
contribution to the $\jj$ absorption
in the hot hadron gas, formed in peripheral heavy-ion collisions
at SPS energy, as discussed thoroughly in Ref.~\cite{ioni2}.
Aiming at calculating thermal averages with
$T\approx 170$~MeV, we didn't discuss in the present paper
the introduction of any arbitrary form factors since the
exponential statistical weight acts as a cut off in the high energy tail.
\section*{Acknowledgments}
 We wish to thank L. Maiani for the stimulating collaboration and encouragement.

\section*{Appendix}
In this Appendix are listed the $I_i$ and $L_i$ integrals
occurring in the calculation and their linear combinations
$A,B,...,F$. These integrals have been computed adopting the
proper time Schwinger regularization prescription, with cut-off
$\mu=0.3$ GeV ($0.5$ GeV when is present a strange quark),
$\Lambda=1.25$~GeV. In the following $N_c=3$.
\begin{eqnarray}
I_1&=&iN_c\int\frac{d^4l}{(2\pi)^4}\frac{1}{(l^2-\tilde
m^2)}\nonumber
\\
&=&\frac{N_c}{16\pi^2}\,\tilde m^2\,\Gamma\left(-1,\frac{\tilde
m^2}{\Lambda^2},\frac{\tilde m^2}{\mu^2}\right)
\\
\nonumber\\
I_3(\delta)&=&-iN_c\int\frac{d^4l}{(2\pi)^4} \frac{1}{(l^2-\tilde
m^2)~(v\cdot l+\delta)}\nonumber
\\
&=&\frac{N_c}{16\pi^{3/2}}\int_{1/{\Lambda^2}}^{1/{\mu^2}}\frac{ds}{s^{3/2}}~
e^{-s\left(\tilde
m^2-\delta^2\right)}~\left(1+{\mathrm{Erf}}\left(\delta\,\sqrt
s\right)\right)
\\
\nonumber\\
I_5(\delta,\delta',\omega)&=&iN_c\int\frac{d^4l}{(2\pi)^4}
\frac{1}{(l^2-\tilde m^2)~(v\cdot l+\delta)~(v'\cdot
l+\delta')}\nonumber
\\
&=&\int_{0}^{1}dy
\frac{1}{1+2y^2(1-\omega)+2y(\omega-1)}\times\nonumber
\\
& &\Bigg[\frac{6}{16\pi^{3/2}}\int_{1/\Lambda^2}^{1/\mu^2}
ds~\sigma~e^{-s\left(\tilde
m^2-\sigma^2\right)}~s^{-1/2}~\left(1+{\mathrm{Erf}}\left(\sigma\,\sqrt
s\right)\right)\nonumber
\\
& &~~~+\frac{6}{16\pi^2}\int_{1/\Lambda^2}^{1/\mu^2} ds~
e^{-s\sigma^2}s^{-1}\Bigg],
\end{eqnarray}
in the last expression we have defined
\begin{equation}
\sigma\equiv\sigma(\delta,\delta',y,\omega)=\frac{\delta\,(1-y)+\delta'\,y}{
\sqrt{1+2\,(\omega-1)\,y+2\,(1-\omega)\,y^2}}.
\end{equation}
In the previous equations $\tilde m^2$, $\delta$ and $\delta'$ are
given by
\begin{eqnarray}
&&\tilde m^2=m^2+x\,m_\rho^2\,(x-1)\\
&&\delta=\Delta-x\,q\cdot v=\Delta-x\,E_\rho\\
&&\delta^\prime=\Delta-x\,q\cdot
v\,\omega=\Delta-x\,\omega\, E_\rho,
\end{eqnarray}
with $m=0.3$ GeV the constituent mass for light quark $u$, $d$.
The expression of $\omega=v\cdot v'$ in the rest frame of $\jj$ is
\begin{equation}
\omega=\frac{m_{\jj}^2+m_\rho^2-m_H^2-m_{H^\prime}^2+2E_\rho
m_{\jj}}{2m_Hm_{H^\prime}}.
\end{equation}

In the $I_1$ integral the gamma-function is
\begin{equation}
\Gamma(\alpha,x_0,x_1) = \int_{x_0}^{x_1} dt\;  e^{-t}\;
t^{\alpha-1},
\end{equation}
while the error function is
\begin{equation}
{\rm Erf}(z)=\frac{2}{\sqrt{\pi}}\int_{0}^{z}dx\ e^{-x^2}.
\end{equation}

The $L_i$ integrals are defined in the following way:
\begin{equation}
L_i=\frac{\partial}{\partial{\tilde m^2}}\,I_i
\end{equation}
and they are
\begin{eqnarray}
L_1&=&\frac{\partial}{\partial{\tilde
m^2}}iN_c\int\frac{d^4l}{(2\pi)^4} \frac{1}{(l^2-\tilde
m^2)}\nonumber
\\
&=&\frac{N_c}{16\pi^2}\left[\Gamma \left(-1,\frac{\tilde
m^2}{\Lambda^2},\frac{\tilde m^2}{\mu^2}\right)+\tilde
m^2\frac{\partial}{\partial{\tilde m^2}}\,
\Gamma\left(-1,\frac{\tilde m^2}{\Lambda^2},\frac{\tilde
m^2}{\mu^2}\right)\right]
\\
\nonumber\\
L_3(\delta)&=&-\frac{\partial}{\partial{\tilde
m^2}}iN_c\int\frac{d^4l}{(2\pi)^4} \frac{1}{(l^2-\tilde
m^2)~(v\cdot l+\delta)}\nonumber
\\
&=&\frac{N_c}{16\pi^{3/2}}\int^{1/\mu^2}_{1/\Lambda^2}ds~
e^{-s\left(\tilde m^2-\delta^2\right)}\left(-s^{-1/2}\right)~
\left(1+{\mathrm{Erf}}\left(\delta\,\sqrt s\right)\right)
\\
\nonumber\\
L_5(\delta,\delta',\omega)&=&\frac{\partial}{\partial{\tilde
m^2}}iN_c\int\frac{d^4l}{(2\pi)^4} \frac{1}{(l^2-\tilde
m^2)~(v\cdot l+\delta)~(v'\cdot l+\delta')}\nonumber
\\
&=&\frac{6}{16\pi^{3/2}}\int^1_0dy\frac{1}{1+2y^2(1-\omega)+2y(\omega-1)}\times\nonumber
\\
& &\times\int^{1/\mu^2}_{1/\Lambda^2}ds~\sigma~ e^{-s\left(\tilde
m^2-\sigma^2\right)}\left(-s^{1/2}\right)~\left(1+{\mathrm{Erf}}\left(\sigma\,\sqrt
s\right)\right).
\end{eqnarray}

The $A,B,C...$ functions are all function of $x$ and $E_\rho$
through $\delta,\delta^\prime$ and $\omega$:
\begin{eqnarray*}
A&=&\frac{L_3(\delta')+\delta'\,
L_5(\delta,\delta',\omega)-(L_3(\delta)+\delta'\,
L_5(\delta,\delta',\omega))\,\omega}{\omega^2-1}
\\
B&=&\frac{L_3(\delta) + \delta'\,
L_5(\delta,\delta',\omega)-(L_3(\delta')+\delta\,
L_5(\delta,\delta',\omega))\,\omega}{\omega^2-1}
\end{eqnarray*}
\begin{eqnarray*}
C&=&\frac{1}{2(\omega^2-1)}\bigg[L_5(\delta,\delta',\omega)\,\delta'^2+(L_3(\delta)-(L_3(\delta')+2\delta
L_5(\delta,\delta',\omega))\,\omega\,\delta'+\delta(L_3(\delta')+\delta
L_5(\delta,\delta',\omega))
\\
&
&-\delta\,L_3(\delta)\,\omega+I_5(\delta,\delta',\omega)\,(\omega^2-1)+L_5(\delta,\delta',\omega)\,
\tilde m^2(\omega^2-1)\bigg]
\\
D&=&\frac{1}{2(\omega^2-1)^2}\bigg[2(L_1+\delta
L_3(\delta))\omega^3+(I_5(\delta,\delta',\omega)+2\delta'(L_3(\delta)+\delta'
L_5(\delta,\delta',\omega))+L_5(\delta,\delta',\omega)\,\tilde
m^2)\omega^2-2L_1\omega-5\delta L_3(\delta)\omega
\\
& &-I_5(\delta,\delta',\omega)+\delta'^2
L_5+3\delta(L_3(\delta')+\delta\,L_5(\delta,\delta',\omega))-L_5(\delta,\delta',\omega)\,\tilde
m^2+\delta'(L_3(\delta)-3(L_3(\delta')+2\delta
L_5(\delta,\delta',\omega))\omega)\bigg]
\\
E&=&\frac{1}{2(\omega^2-1)^2}\bigg[2(L_1+\delta'L_3(\delta'))\omega^3+(I_5(\delta,\delta',\omega)
+2\delta(L_3(\delta')+\delta
L_5(\delta,\delta',\omega))+L_5(\delta,\delta',\omega)\,\tilde
m^2)\omega^2-(2L_1+5\delta'L_3(\delta')
\\
&
&+3\delta(L_3(\delta)+2\delta'L_5(\delta,\delta',\omega)))\omega-I_5(\delta,\delta',\omega)+3\delta'L_3(\delta)+\delta
L_3(\delta')+\delta^2\,L_5(\delta,\delta',\omega)+3\delta'^2\,L_5(\delta,\delta',\omega)-L_5(\delta,\delta',\omega)
\tilde m^2\bigg]
\\
F&=&\frac{1}{2(\omega^2 -
1)^2}\bigg[(-(I_5(\delta,\delta',\omega)+L_5(\delta,\delta',\omega)\,\tilde
m^2))\omega^3+(-2L_1+\delta'L_3(\delta')+\delta(L_3(\delta)+4\delta'L_5(\delta,\delta',\omega)))\omega^2
\\
& &+(I_5(\delta,\delta',\omega)-3(\delta(L_3(\delta)+\delta
L_5(\delta,\delta',\omega))+\delta'(L_3(\delta)+\delta'L_5(\delta,\delta',\omega)))+L_5(\delta,\delta',\omega)\,\tilde
m^2)\omega
\\
& &+2L_1+2(\delta'L_3(\delta)
+\delta(L_3(\delta)+\delta'L_5(\delta,\delta',\omega)))\bigg].
\end{eqnarray*}

The determined couplings weight the scalar combinations of
external particle momenta and polarizations in the combinations
obtained by the loop computation in Eq.~(\ref{eq.VMD}). We list
below the scalar combinations $S\left(D^{(*)}D^{(*)}\right)$
together with the couplings they are weighted by, and with the
effective Lagrangian interactions ${\cal L}$ one can write down
for them. These terms describe the four-linear diagrams in
Fig.~1; for the actual cross section computation one has to
add also the $t$ and $u$ channels.

\begin{displaymath}
\begin{array}{crc}
\hline\hline
~~~~~~~~~~~~~~~S(DD)~~~~~~~~~~~~~~~&g&~~~~~~~~~~~~~~~~~~~~{\cal L}~~~~~~~~~~~~~~~~~~~~\\
\hline
\\
q\cdot p_2~\eta\cdot p_1~\epsilon\cdot p_1&-g_1&
-J_\alpha~\partial_\mu\rho_\nu~\partial^\mu D~\partial^\nu\partial^\alpha\bar D\\
q\cdot p_2~\eta\cdot p_2~\epsilon\cdot p_1&g_2&
J_\alpha~\partial_\mu\rho_\nu~\partial^\mu\partial^\alpha D~\partial^\nu\bar D\\
q\cdot p_1~\epsilon\cdot p_2~\eta\cdot p_1&g_1&
-J_\alpha~\partial_\mu\rho_\nu~\partial^\mu\partial^\alpha\bar D~\partial^\nu D\\
q\cdot p_1~\epsilon\cdot p_2~\eta\cdot p_2&-g_2&
J_\alpha~\partial_\mu\rho_\nu~\partial^\mu\bar D~\partial^\nu\partial^\alpha D\\
\eta\cdot q~\epsilon\cdot p_1&g_3&
-\partial_\mu\rho_\nu~J^\mu~\partial^\nu\bar D~ D\\
\eta\cdot p_1~\epsilon\cdot p_1&g_4&
-J_\mu\rho_\nu\partial^\mu\partial^\nu\bar D~D\\
\eta\cdot p_2~\varepsilon\cdot p_1&g_5&
J\cdot\partial D~\rho\cdot\partial\bar D\\
q\cdot\eta~\epsilon\cdot p_2&-g_6&
\partial_\mu\rho_\nu~J^\mu~\partial^\nu D~\bar D\\
q\cdot p_1~\epsilon\cdot\eta&-g_3&
-\partial_\mu\rho_\nu~J^\nu~\partial^\mu\bar D~D\\
q\cdot p_2~\epsilon\cdot\eta&g_6&
\partial_\mu\rho_\nu~J^\nu~\partial^\mu D~\bar D\\
\epsilon\cdot p_2~\eta\cdot p_1&g_7&
J\cdot\partial\bar D~\rho\cdot\partial D\\
\epsilon\cdot p_2~\eta\cdot p_2&g_8&
-J_\mu\rho_\nu\partial^\mu\partial^\nu D~\bar D\\
\eta\cdot\epsilon&g_9&J\cdot\rho~D~\bar D\\
\\
\hline\hline
\end{array}
\end{displaymath}

\begin{displaymath}
\begin{array}{ccc}
\hline\hline
~~~~~~~~~~~~~~~S(DD^*)~~~~~~~~~~~~~~~&h&~~~~~~~~~~~~~~~~~~~~{\cal L}~~~~~~~~~~~~~~~~~~~~\\
\hline
\\
i\varepsilon_{\alpha\beta\gamma\delta}~q^\alpha\epsilon^\beta\eta^\gamma\eta_1^\delta&h_1
&-\varepsilon_{\alpha\beta\gamma\delta}~\partial^\alpha\rho^\beta
J^\gamma\bar {D^*}^\delta D
\\
i\varepsilon_{\alpha\beta\gamma\delta}~q^\alpha\epsilon^\beta\eta^\gamma
p_1^\delta~~\eta_1\cdot p_2&\frac{1}{m_{D^*}m_D}h_2
&-\varepsilon_{\alpha\beta\gamma\delta}~\partial^\alpha\rho^\beta
J^\gamma\partial^\delta\bar{D^*}^\mu\partial_\mu D
\\
i\varepsilon_{\alpha\beta\gamma\delta}~q^\alpha\epsilon^\beta\eta^\gamma
p_2^\delta~~\eta_1\cdot p_2&-\frac{1}{m^2_D}h_2
&\varepsilon_{\alpha\beta\gamma\delta}~\partial^\alpha\rho^\beta
J^\gamma\partial^\delta\partial_\mu D \bar{D^*}^\mu
\\
i\varepsilon_{\alpha\beta\gamma\delta}~q^\alpha\epsilon^\beta\eta_1^\gamma
p_1^\delta~~\eta\cdot p_1&-\frac{1}{m^2_{D^*}}(h_5+h_6)
&\varepsilon_{\alpha\beta\gamma\delta}~\partial^\alpha\rho^\beta\partial^\delta\partial^\mu
\bar{D^*}^\gamma J_\mu D
\\
i\varepsilon_{\alpha\beta\gamma\delta}~q^\alpha\epsilon^\beta\eta_1^\gamma
p_1^\delta~~\eta\cdot p_2&\frac{1}{m_{D^*}m_D}h_3 &
\varepsilon_{\alpha\beta\gamma\delta}~\partial^\alpha\rho^\beta\partial^\delta\bar{D^*}^\gamma
J\cdot\partial D
\\
i\varepsilon_{\alpha\beta\gamma\delta}~q^\alpha\epsilon^\beta\eta_1^\gamma
p_2^\delta~~\eta\cdot p_1&\frac{1}{m_{D^*}m_D}h_4
&-\varepsilon_{\alpha\beta\gamma\delta}~\partial^\alpha\rho^\beta\partial^\mu\bar{D^*}^\gamma
\partial^\delta DJ_\mu
\\
i\varepsilon_{\alpha\beta\gamma\delta}~q^\alpha\epsilon^\beta\eta_1^\gamma
p_2^\delta~~\eta\cdot p_2&\frac{1}{m^2_D}h_2
&\varepsilon_{\alpha\beta\gamma\delta}~\partial^\alpha\rho^\beta\bar{D^*}^\gamma\partial^\delta\partial_\mu
DJ^\mu
\\
i\varepsilon_{\alpha\beta\gamma\delta}~q^\alpha\epsilon^\beta
p_1^\gamma p_2^\delta~~\eta\cdot\eta_1&\frac{1}{m_{D^*}m_D}h_5
&-\varepsilon_{\alpha\beta\gamma\delta}~\partial^\alpha\rho^\beta\partial^\gamma\bar{D^*}^\mu
\partial^\delta D J_\mu
\\
i\varepsilon_{\alpha\beta\gamma\delta}~q^\alpha\eta^\beta\eta_1^\gamma
p_1^\delta~~\epsilon\cdot p_1&\frac{1}{m^2_{D^*}}(h_5-h_6)
&\varepsilon_{\alpha\beta\gamma\delta}~\partial^\alpha\rho^\mu
J^\beta\partial^\delta\partial_\mu\bar{D^*}^\gamma D
\\
i\varepsilon_{\alpha\beta\gamma\delta}~q^\alpha\eta^\beta\eta_1^\gamma
p_1^\delta~~\epsilon\cdot p_2&\frac{1}{m_{D^*}m_D}(-h_5-h_2)
&-\varepsilon_{\alpha\beta\gamma\delta}~\partial^\alpha\rho^\mu
J^\beta\partial^\delta\bar{D^*}^\gamma\partial_\mu D
\\
i\varepsilon_{\alpha\beta\gamma\delta}~q^\alpha\eta^\beta\eta_1^\gamma
p_2^\delta~~\epsilon\cdot p_1&\frac{1}{m_{D^*}m_D}h_6
&-\varepsilon_{\alpha\beta\gamma\delta}~\partial^\alpha\rho^\mu
J^\beta\partial_\mu\bar{D^*}^\gamma\partial^\delta D
\\
i\varepsilon_{\alpha\beta\gamma\delta}~q^\alpha\eta^\beta\eta_1^\gamma
p_2^\delta~~\epsilon\cdot p_2&\frac{1}{m^2_D}h_2
&\varepsilon_{\alpha\beta\gamma\delta}~\partial^\alpha\rho^\mu
J^\beta\bar{D^*}^\gamma\partial^\delta\partial_\mu D
\\
i\varepsilon_{\alpha\beta\gamma\delta}~q^\alpha\eta^\beta
p_1^\gamma p_2^\delta~~\epsilon\cdot\eta_1&\frac{1}{m_{D^*}m_D}h_7
&-\varepsilon_{\alpha\beta\gamma\delta}~\partial^\alpha\rho^\mu
J^\beta\partial^\gamma\bar{D^*}_\mu\partial^\delta D
\\
i\varepsilon_{\alpha\beta\gamma\delta}~\epsilon^\alpha\eta^\beta
\eta_1^\gamma p_1^\delta~~q\cdot p_1&-\frac{1}{m^2_{D^*}}(h_5-h_6)
&\varepsilon_{\alpha\beta\gamma\delta}~\partial^\mu\rho^\alpha
J^\beta\partial^\delta\partial_\mu\bar{D^*}^\gamma D
\\
i\varepsilon_{\alpha\beta\gamma\delta}~\epsilon^\alpha\eta^\beta
\eta_1^\gamma p_1^\delta~~q\cdot p_2&\frac{1}{m_{D^*}m_D}(h_5-h_6)
&-\varepsilon_{\alpha\beta\gamma\delta}~\partial^\mu\rho^\alpha
J^\beta\partial^\delta\bar{D^*}^\gamma\partial_\mu D
\\
i\varepsilon_{\alpha\beta\gamma\delta}~\epsilon^\alpha\eta^\beta
\eta_1^\gamma p_1^\delta&\frac{1}{m_{D^*}}h_8
&-\varepsilon_{\alpha\beta\gamma\delta}~\rho^\alpha
J^\beta\partial^\delta\bar{D^*}^\gamma D
\\
i\varepsilon_{\alpha\beta\gamma\delta}~\epsilon^\alpha\eta^\beta
\eta_1^\gamma p_2^\delta~~q\cdot p_1&\frac{1}{m_{D^*}m_D}h_2
&-\varepsilon_{\alpha\beta\gamma\delta}~\partial^\mu\rho^\alpha
J^\beta\partial_\mu\bar{D^*}^\gamma\partial^\delta D
\\
i\varepsilon_{\alpha\beta\gamma\delta}~\epsilon^\alpha\eta^\beta
\eta_1^\gamma p_2^\delta~~q\cdot p_2&-\frac{1}{m^2_D}h_2
&\varepsilon_{\alpha\beta\gamma\delta}~\partial^\mu\rho^\alpha
J^\beta\bar{D^*}^\gamma\partial^\delta\partial_\mu D
\\
i\varepsilon_{\alpha\beta\gamma\delta}~\epsilon^\alpha\eta^\beta
\eta_1^\gamma p_2^\delta&-\frac{1}{m_D}h_8
&-\varepsilon_{\alpha\beta\gamma\delta}~\rho^\alpha
J^\beta\bar{D^*}^\gamma\partial^\delta D
\\
i\varepsilon_{\alpha\beta\gamma\delta}~\epsilon^\alpha\eta^\beta
p_1^\gamma p_2^\delta~~q\cdot\eta_1&\frac{1}{m_{D^*}m_D}(h_3-h_6)
&-\varepsilon_{\alpha\beta\gamma\delta}~\partial^\mu\rho^\alpha
J^\beta\partial^\gamma\bar{D^*}_\mu\partial^\delta D
\\
i\varepsilon_{\alpha\beta\gamma\delta}~\epsilon^\alpha\eta_1^\beta
p_1^\gamma p_2^\delta~~q\cdot\eta&\frac{1}{m_{D^*}m_D}(h_2+h_6)
&-\varepsilon_{\alpha\beta\gamma\delta}~\partial^\mu\rho^\alpha\partial^\gamma\bar{D^*}^\beta
\partial^\delta DJ_\mu
\\
i\varepsilon_{\alpha\beta\gamma\delta}~\eta^\alpha\eta_1^\beta
p_1^\gamma p_2^\delta~~\epsilon\cdot
p_1&\frac{-2}{m^2_{D^*}m_D}h_9
&-\varepsilon_{\alpha\beta\gamma\delta}~J^\alpha\partial^\gamma\partial^\mu\bar{D^*}^\beta
\partial^\delta D\rho_\mu
\\
i\varepsilon_{\alpha\beta\gamma\delta}~\eta^\alpha\eta_1^\beta
p_1^\gamma p_2^\delta~~\epsilon\cdot
p_2&\frac{-2}{m_{D^*}m^2_D}h_{10}
&\varepsilon_{\alpha\beta\gamma\delta}~J^\alpha\partial^\gamma\bar{D^*}^\beta\partial^\delta\partial_\mu
D\rho^\mu
\\
\\
\hline\hline
\end{array}
\end{displaymath}

\begin{displaymath}
\begin{array}{crc}
\hline\hline
~~~~~~~~~~~~~~~S(D^*D^*)~~~~~~~~~~~~~~~&f&~~~~~~~~~~~~~~~~~~~~{\cal L}~~~~~~~~~~~~~~~~~~~~\\
\hline
\\
q\cdot\eta~\epsilon\cdot\eta_1~\eta_2\cdot
p_1&f_1&-\partial_\mu\rho_\nu
J^\mu\partial_\alpha\bar{D^*}^\nu{D^*}^\alpha
\\
q.\eta~\epsilon.\eta_2~\eta_1\cdot p_2&-f_1&\partial_\mu\rho_\nu
J^\mu\partial_\alpha{D^*}^\nu\bar{D^*}^\alpha
\\
q\cdot \eta~\epsilon\cdot p_1~\eta_1\cdot
\eta_2&-f_2&-\partial_\mu\rho_\nu
J^\mu\partial^\nu\bar{D^*}_\alpha{D^*}^\alpha
\\
q\cdot \eta~\epsilon\cdot p_1~\eta_1\cdot p_2~\eta_2\cdot
p_1&f_3&-\partial_\mu\rho_\nu
J^\mu\partial^\nu\partial_\beta\bar{D^*}^\alpha\partial_\alpha{D^*}^\beta
\\
q\cdot \eta~\epsilon\cdot p_2~\eta_1\cdot
\eta_2&f_4&\partial_\mu\rho_\nu
J^\mu\partial^\nu{D^*}_\alpha\bar{D^*}^\alpha
\\
q\cdot \eta~\epsilon\cdot p_2~\eta_1\cdot p_2~\eta_2\cdot
p_1&f_5&\partial_\mu\rho_\nu
J^\mu\partial^\nu\partial_\alpha{D^*}_\beta\partial^\beta\bar{D^*}^\alpha
\\
q\cdot \eta_1~\epsilon\cdot \eta~\eta_2\cdot
p_1&-f_1&-\partial_\mu\rho_\nu
J^\nu\partial_\alpha\bar{D^*}^\mu{D^*}^\alpha
\\
q\cdot \eta_1~\epsilon\cdot \eta_2~\eta\cdot
p_1&f_1&-\partial_\mu\rho_\nu\partial^\mu\bar{D^*}_\alpha
J^\alpha{D^*}^\nu
\\
q\cdot \eta_1~\epsilon\cdot \eta_2~\eta\cdot
p_2&f_1&\partial_\mu\rho_\nu\bar{D^*}^\mu\partial_\alpha{D^*}^\nu
J^\alpha
\\
q\cdot \eta_1~\epsilon\cdot p_1~\eta\cdot
\eta_2&f_2&-\partial_\mu\rho_\nu\partial^\nu\bar{D^*}^\mu J\cdot
{D^*}
\\
q\cdot \eta_1~\epsilon\cdot p_1~\eta\cdot p_2~\eta_2\cdot
p_1&-f_3&-\partial_\mu\rho_\nu\partial^\nu\partial_\beta\bar{D^*}^\mu
J_\alpha\partial^\alpha{D^*}^\beta
\\
q\cdot \eta_1~\epsilon\cdot p_2~\eta\cdot
\eta_2&-f_4&\partial_\mu\rho_\nu\bar{D^*}^\mu
J_\alpha\partial^\nu{D^*}^\alpha
\\
q\cdot \eta_1~\epsilon\cdot p_2~\eta\cdot p_2~\eta_2\cdot
p_1&-f_5&\partial_\mu\rho_\nu\partial_\alpha\bar
{D^*}^\mu\partial^\nu\partial_\beta{D^*}^\alpha J^\beta
\\
q\cdot \eta_2~\epsilon\cdot \eta~\eta_1\cdot
p_2&f_1&\partial_\mu\rho_\nu\bar{D^*}_\alpha\partial^\alpha{D^*}^\mu
J^\nu
\\
q\cdot \eta_2~\epsilon\cdot \eta_1~\eta\cdot
p_1&-f_1&-\partial_\mu\rho_\nu\partial_\alpha\bar{D^*}^\nu{D^*}^\mu
J^\alpha
\\
q\cdot \eta_2~\epsilon\cdot \eta_1~\eta\cdot
p_2&-f_1&\partial_\mu\rho_\nu\bar{D^*}^\nu\partial_\alpha{D^*}^\mu
J_\alpha
\\
q\cdot \eta_2~\epsilon\cdot p_1~\eta\cdot
\eta_1&f_2&-\partial_\mu\rho_\nu\partial^\nu\bar{D^*}_\alpha{D^*}^\mu
J_\alpha
\\
q\cdot \eta_2~\epsilon\cdot p_1~\eta\cdot p_1~\eta_1\cdot
p_2&-f_3&-\partial_\mu\rho_\nu\partial^\nu\partial_\alpha\bar{D^*}_\beta
\partial^\beta{D^*}^\mu J^\alpha
\\
q\cdot \eta_2~\epsilon\cdot p_2~\eta\cdot
\eta_1&-f_4&\partial_\mu\rho_\nu\bar{D^*}_\alpha\partial^\nu{D^*}^\mu
J^\alpha
\\
q\cdot \eta_2~\epsilon\cdot p_2~\eta\cdot p_1~\eta_1\cdot
p_2&-f_5&\partial_\mu\rho_\nu\partial_\alpha\bar{D^*}_\beta
\partial^\nu\partial^\beta{D^*}^\mu
J^\alpha
\\
q\cdot p_1~\epsilon\cdot \eta~\eta_1\cdot
\eta_2&f_2&-\partial_\mu\rho_\nu\partial^\mu\bar{D^*}_\alpha{D^*}^\alpha
J^\nu
\\
q\cdot p_1~\epsilon\cdot \eta~\eta_1\cdot p_2~\eta_2\cdot
p_1&-f_3&-\partial_\mu\rho_\nu\partial^\mu\partial_\beta\bar{D^*}_\alpha
\partial^\alpha{D^*}^\beta J^\nu
\\
q\cdot p_1~\epsilon\cdot \eta_1~\eta\cdot
\eta_2&-f_2&-\partial_\mu\rho_\nu\partial^\mu\bar{D^*}^\nu{D^*}_\alpha
J^\alpha
\\
q\cdot p_1~\epsilon\cdot \eta_1~\eta\cdot p_2~\eta_2\cdot
p_1&f_3&-\partial_\mu\rho_\nu\partial^\mu\partial_\beta\bar{D^*}^\nu
\partial_\alpha{D^*}^\beta J^\alpha
\\
q\cdot p_1~\epsilon\cdot \eta_2~\eta\cdot
\eta_1&-f_2&-\partial_\mu\rho_\nu\partial^\mu\bar{D^*}_\alpha{D^*}^\nu
J^\alpha
\\
q\cdot p_1~\epsilon\cdot \eta_2~\eta\cdot p_1~\eta_1\cdot
p_2&f_3&\partial_\mu\rho_\nu\partial^\mu\partial_\alpha\bar{D^*}_\beta
\partial^\beta{D^*}^\nu J^\alpha
\\
q\cdot p_1~\epsilon\cdot p_2~\eta\cdot \eta_1~\eta_2\cdot
p_1&f_3&-\partial_\mu\rho_\nu\partial^\mu\partial_\beta\bar{D^*}_\alpha
\partial^\nu{D^*}^\beta J^\alpha
\\
q\cdot p_1~\epsilon\cdot p_2~\eta\cdot \eta_2~\eta_1\cdot
p_2&-f_5&\partial_\mu\rho_\nu\partial^\mu\bar{D^*}_\beta
\partial^\nu\partial^\beta{D^*}_\alpha J^\alpha
\\
q\cdot p_1~\epsilon\cdot p_2~\eta\cdot p_1~\eta_1\cdot
\eta_2&-f_3&-\partial_\mu\rho_\nu\partial^\mu\partial_\alpha\bar{D^*}_\beta
\partial^\nu{D^*}^\beta J^\alpha
\\
q\cdot p_1~\epsilon\cdot p_2~\eta\cdot p_2~\eta_1\cdot
\eta_2&f_5&\partial_\mu\rho_\nu\partial^\mu\bar{D^*}_\beta
\partial^\nu\partial_\alpha{D^*}^\beta J^\alpha
\\
q\cdot p_2~\epsilon\cdot \eta\eta_1\cdot
\eta_2&-f_4&\partial_\mu\rho_\nu\bar{D^*}_\alpha\partial^\mu{D^*}^\alpha
J^\nu
\\
q\cdot p_2\epsilon\cdot \eta~\eta_1\cdot p_2~\eta_2\cdot
p_1&-f_5&\partial_\mu\rho_\nu\partial_\beta\bar{D^*}_\alpha
\partial^\mu\partial^\alpha{D^*}^\beta J^\nu
\\
q\cdot p_2~\epsilon\cdot \eta_1~\eta\cdot
\eta_2&f_4&\partial_\mu\rho_\nu\bar{D^*}^\nu\partial^\mu{D^*}_\alpha
J^\alpha
\\
q\cdot p_2~\epsilon\cdot \eta_1~\eta\cdot p_2~\eta_2\cdot
p_1&f_5&\partial_\mu\rho_\nu\partial_\beta\bar{D^*}^\nu
\partial^\mu\partial_\alpha{D^*}^\beta J^\alpha
\\
q\cdot p_2~\epsilon\cdot \eta_2~\eta\cdot
\eta_1&f_4&\partial_\mu\rho_\nu\bar{D^*}_\alpha\partial^\mu{D^*}^\nu
J^\alpha
\\
q\cdot p_2~\epsilon\cdot \eta_2~\eta\cdot p_1~\eta_1\cdot
p_2&f_5&\partial_\mu\rho_\nu\partial_\alpha\bar{D^*}_\beta
\partial^\mu\partial^\beta{D^*}^\nu J^\alpha
\\
q\cdot p_2~\epsilon\cdot p_1~\eta\cdot \eta_1~\eta_2\cdot
p_1&-f_3&-\partial_\mu\rho_\nu\partial^\nu\partial_\beta\bar{D^*}_\alpha
\partial^\mu{D^*}^\beta J^\alpha
\\
q\cdot p_2~\epsilon\cdot p_1~\eta\cdot \eta_2~\eta_1\cdot
p_2&f_5&\partial_\mu\rho_\nu\partial^\nu\bar{D^*}_\beta
\partial^\mu\partial^\beta{D^*}_\alpha J^\alpha
\\
q\cdot p_2~\epsilon\cdot p_1~\eta\cdot p_1~\eta_1\cdot
\eta_2&f_3&-\partial_\mu\rho_\nu\partial^\nu\partial_\alpha\bar{D^*}_\beta
\partial^\mu{D^*}^\beta J^\alpha
\\
q\cdot p_2~\epsilon\cdot p_1~\eta\cdot p_2~\eta_1\cdot
\eta_2&-f_5&\partial_\mu\rho_\nu\partial^\nu\bar{D^*}_\beta
\partial^\mu\partial_\alpha{D^*}^\beta J^\alpha
\\
\epsilon\cdot \eta~\eta_1\cdot \eta_2&f_6&\rho\cdot
J\bar{D^*}\cdot {D^*}
\\
\epsilon\cdot \eta~\eta_1\cdot p_2~\eta_2\cdot p_1&f_7&\rho\cdot
J\partial_\alpha\bar{D^*}_\beta\partial^\beta{D^*}^\alpha
\\
\epsilon\cdot \eta_1~\eta\cdot \eta_2&-f_6&\rho\cdot
\bar{D^*}J\cdot {D^*}
\\
\epsilon\cdot \eta_1~\eta\cdot p_2~\eta_2\cdot
p_1&-f_7&\rho_\alpha\partial_\gamma\bar{D^*}^\alpha\partial_\beta{D^*}^\gamma
J^\beta
\\
\epsilon\cdot \eta_2~\eta\cdot \eta_1&-f_6&\rho\cdot {D^*}J\cdot
\bar{D^*}
\\
\epsilon\cdot \eta_2~\eta\cdot p_1~\eta_1\cdot
p_2&-f_7&\rho_\alpha\partial_\beta\bar{D^*}_\gamma\partial^\gamma{D^*}^\alpha
J^\beta
\\
\epsilon\cdot p_1~\eta\cdot \eta_1~\eta_2\cdot
p_1&f_8&-\rho_\alpha\partial^\alpha\partial_\gamma\bar{D^*}_\beta{D^*}^\gamma
J^\beta
\\
\epsilon\cdot p_1~\eta\cdot \eta_2~\eta_1\cdot
p_2&-f_9&\rho_\alpha\partial^\mu\bar{D^*}_\gamma\partial^\gamma{D^*}_\beta
J^\beta
\\
\epsilon\cdot p_1~\eta\cdot p_1~\eta_1\cdot
\eta_2&-f_8&-\rho_\alpha\partial^\alpha\partial_\beta\bar{D^*}_\gamma{D^*}^\gamma
J^\beta
\\
\epsilon\cdot p_1~\eta\cdot p_2~\eta_1\cdot
\eta_2&f_9&\rho_\alpha\partial^\alpha\bar{D^*}_\gamma\partial_\beta{D^*}^\gamma
J^\beta
\\
\epsilon\cdot p_2~\eta\cdot \eta_1~\eta_2\cdot
p_1&-f_{10}&\rho_\alpha\partial_\gamma\bar{D^*}_\beta\partial^\alpha{D^*}^\gamma
J^\beta
\\
\epsilon\cdot p_2~\eta\cdot \eta_2~\eta_1\cdot
p_2&f_{11}&-\rho_\alpha\bar{D^*}_\gamma\partial^\alpha\partial^\gamma{D^*}_\beta
J^\beta
\\
\epsilon\cdot p_2~\eta\cdot p_1~\eta_1\cdot
\eta_2&f_{10}&\rho_\alpha\partial_\beta\bar{D^*}_\gamma\partial^\alpha{D^*}^\gamma
J^\beta
\\
\epsilon\cdot p_2~\eta\cdot p_2~\eta_1\cdot
\eta_2&-f_{11}&-\rho_\alpha\bar{D^*}_\gamma\partial^\alpha\partial_\beta{D^*}^\gamma
J^\beta
\\
\\
\hline\hline
\end{array}
\end{displaymath}

\end{document}